\documentclass[
 aip,
 rsi,
 amsmath,amssymb,
 reprint,%
]{revtex4-1}

\usepackage{graphicx}
\usepackage{dcolumn}
\usepackage{bm}

\usepackage[utf8]{inputenc}
\usepackage[T1]{fontenc}
\usepackage{mathptmx}
\usepackage{etoolbox}
\usepackage{braket}

\makeatletter
\def\@email#1#2{%
 \endgroup
 \patchcmd{\titleblock@produce}
  {\frontmatter@RRAPformat}
  {\frontmatter@RRAPformat{\produce@RRAP{*#1\href{mailto:#2}{#2}}}\frontmatter@RRAPformat}
  {}{}
}%
\makeatother
\begin{document}

\preprint{AIP/123-QED}

\title[Giant spontaneous polarization in zincblende III-V semiconductors]{Giant spontaneous polarization in zincblende III-V semiconductors\\}
\author{J. Cañas}

\email{jesus.canas@neel.cnrs.fr; mohamed.yassine@inatech.uni-freiburg.de}

\affiliation{Power Electronics, Institute for Sustainable Systems Engineering INATECH, University of Freiburg, Emmy-Noether-Str. 2,
D-79110 Freiburg, Germany}%
\affiliation{Univ. Grenoble Alpes, CNRS, Grenoble INP, Institut Néel, 38000 Grenoble, France}%

\author{M. Yassine}

\affiliation{Power Electronics, Institute for Sustainable Systems Engineering INATECH, University of Freiburg, Emmy-Noether-Str. 2,
D-79110 Freiburg, Germany}

\author{O. Ambacher}
\affiliation{Power Electronics, Institute for Sustainable Systems Engineering INATECH, University of Freiburg, Emmy-Noether-Str. 2,
D-79110 Freiburg, Germany}

\date{\today}

\begin{abstract}

The discovery of ferroelectricity in wurtzite nitrides has paved the way for measuring and understanding spontaneous polarization in III-V semiconductors. However, the calculation of polarization effects at heterointerfaces — crucial for numerous electronic and photonic applications — remains a topic of debate. The need for a reference structure to calculate spontaneous polarization has led to discussions over whether to use the zincblende or layered hexagonal structures as the reference for wurtzite crystals. In this work, we argue that the layered hexagonal structure is not only a better reference due to its vanishing formal polarization but also the only physically correct choice for the wurtzite system. This follows from the fact that spontaneous polarization is rigorously defined through the ferroelectric switching. Applying this definition, we extend our analysis to III-V zincblende semiconductors and reveal that their spontaneous polarization is approximately three times larger than that of wurtzite, thereby refuting the longstanding assumption that it is zero. Through this example, we illustrate that spontaneous polarization is not inherently linked to charge density at interfaces.

\end{abstract}

\maketitle

The recent measurement of ferroelectricity in wurtzite ScAlN has established a basis for understanding spontaneous polarization in III-V semiconductors.\cite{Fichtner2019}  The need for a reference structure to calculate spontaneous polarization has led to a debate over whether to use the zincblende \cite{Bernardini} or layered hexagonal \cite{Dreyer2016} structures as the reference for wurtzite crystals. Due to this discrepancy, the polarization effects at heterointerfaces — critical for key applications such as light-emitting diodes (LEDs) and high-electron-mobility transistors (HEMTs) — remain a subject of discussion. The choice of reference significantly impacts the deduced values of spontaneous polarization, differing by orders of magnitude. However, predicted values for polarization-bound charges at conventional nitride interfaces show only slight variations. Notably, when stacking different crystal types, giant polarization discontinuities have been predicted using the layered hexagonal reference.\cite{Adamski2019,Adamski2020-oxides} This striking yet experimentally unverified prediction\cite{Dinh-electrical-ScN} has led much of the nitride research community to rely on spontaneous polarization values derived from the zincblende reference.

 In this contribution, we resolve this discrepancy by revisiting the unequivocal definition of spontaneous polarization, tied to ferroelectric switching, that connects spontaneous polarization to a physically measurable quantity.\cite{kanzig1957ferroelectrics,resta2007theory} Indeed, the measured values of spontaneous polarization from ferroelectric measurements in wurtzite nitrides align with calculations that use the layered hexagonal structure as a reference. This is because the layered hexagonal structure represents an intermediate centrosymmetric phase between the polarities of wurtzite.\cite{Mo2024} Moreover, we extend this definition to study the III-V zincblende semiconductor system. We show that zincblende semiconductors exhibit giant spontaneous polarization up to three times that of wurtzite crystals, contrary to the historical belief that it has no spontaneous polarization. Through this example, we demonstrate that spontaneous polarization is not necessarily related to polarization-bound charges at interfaces.

Under the modern theory of polarization, the formal polarization ($P_f$) of a crystal is determined as the sum of two components: ionic polarization, which arises from the contribution of atomic nuclei and is computed using the classical description of dipoles, and electronic polarization, which originates from the valence electrons and is calculated using the Berry phase approach:\cite{Vanderbilt1993,Resta1994,resta2007theory}

\begin{equation}
\begin{aligned}
P_f= \frac{e}{\Omega} \sum_s Z_s^{ion} R_s^{(\lambda)} + \frac{i e\Omega}{(2\pi)^3} \sum_j^{occ.} \int_{BZ} dk \braket{u_{j,k}^{(\lambda)}|\Delta_k|u_{j,k}^{(\lambda)}}
\end{aligned}
\end{equation}

The first term is the ionic contribution ($P_{ion}$), calculated as the sum of the product of the ionic charges $Z_s$ and the nuclei positions $R_s$ over the unit cell volume $\Omega$. The second term is the electronic contribution ($P_{el}$), calculated through the Berry potential $\braket{u_{j,k}^{(\lambda)}|\Delta_k|u_{j,k}^{(\lambda)}}$ of the cell-periodic crystal Bloch wavefunctions $\ket{u_{j,k}^{(\lambda)}}$ integrated over the Brillouin zone. This second term represents the electronic polarization as the sum of the product of the electronic charge and the Wannier centers. 

\begin{figure*}
\includegraphics[width=13cm]{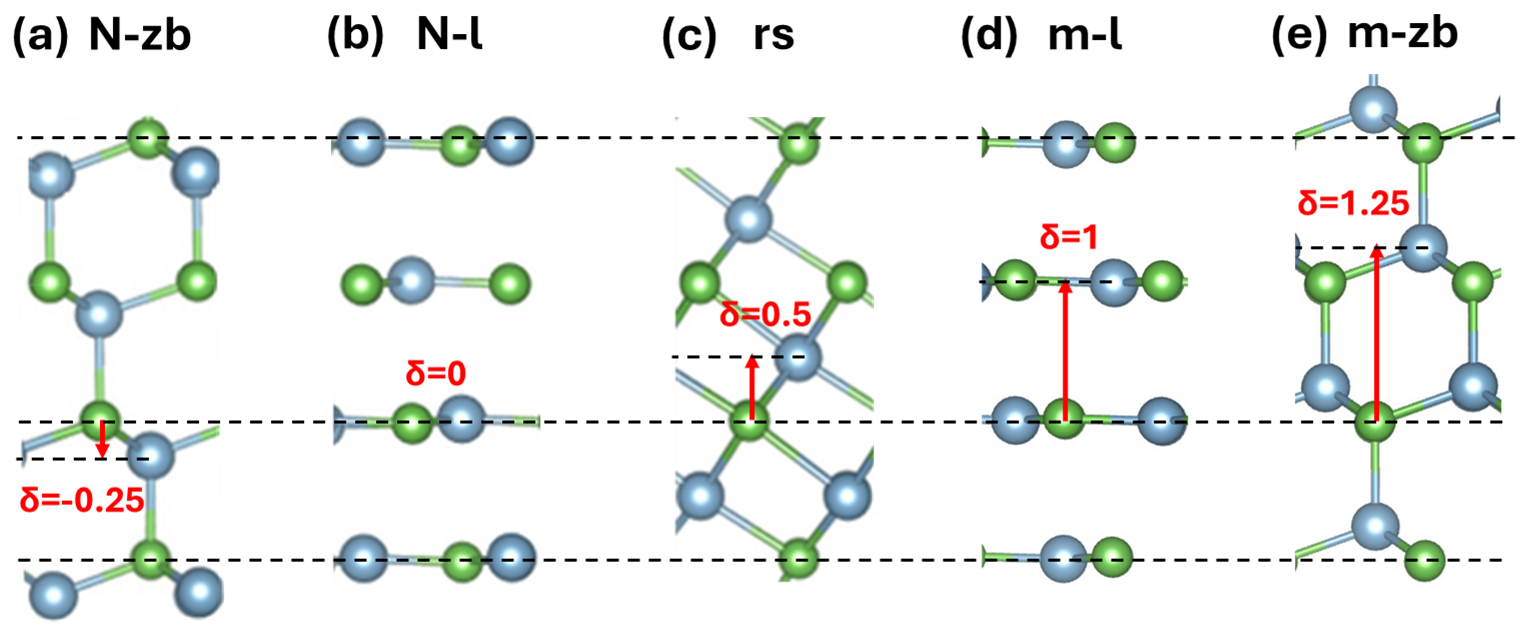}

\caption{Scheme of the zincblende N-polar to m-polar transformation along the <111> direction showing the ratio of the separation between planes of anions (green) and cations (blue) to the separation between planes of cations in each structure ($\delta$). (a) nitrogen-polar zincblende, (b) nitrogen-polar layered, (c) rocksalt, (d) metal-polar layered and (e) metal-polar zincblende.}
\label{schem}
\end{figure*}

The formal polarization corresponds to the dipolar moment of a primitive unit cell divided by its volume. As a result, the formal polarization depends on the location (and the shape) of the primitive unit cell. As there are multiple choices of primitive cell to describe a crystal, the polarization of an infinite crystal becomes a multivalued quantity—forming a polarization lattice rather than a single value as explained by Spaldin.\cite{spaldin2012} This multivalued nature emerges directly from the crystal's periodicity and the invariance under unit cell translation.\cite{spaldin2012,Multivalueness,resta2007theory} The multivalued character of polarization gives rise to the polarization quantum ($P_q$), which separates the different values (branches) of polarization. $P_q$ is given by $eR/\Omega$, where R is the vector along which the polarization is calculated.  Consequently, the polarization of a crystal is not uniquely defined, and only differences in formal polarization (within the correct polarization branch) have physical significance. Indeed, this is also the case for experimental measurements of polarization, as all polarization-induced effects (such as interface charges, ferroelectric switching, or piezoelectric effect) are always observed through polarization differences.

The definition of spontaneous polarization has been a longstanding challenge within the nitride community.  
Bernardini proposed defining the spontaneous polarization of wurtzite by taking its formal polarization difference with a zincblende reference.\cite{Bernardini} The intuition behind this approach is that the spontaneous polarization of zincblende is zero, and both wurtzite and zincblende structures are tetrahedrally coordinated. Therefore, any structural deviations from the ideal wurtzite structure would lead to a formal polarization difference relative to its zincblende counterpart, giving rise to spontaneous polarization. While the definition of such a formal polarization difference is theoretically sound, referring to it as spontaneous polarization is somewhat arbitrary, as other reference structures could be chosen ambiguously. Moreover, there is no straightforward transformation from a wurtzite to a zincblende structure. \cite{spaldin2012,resta2007theory, Bernardini} The lack of ferroelectricity in these materials until recent discoveries \cite{Fichtner2019} created a disconnect between the nitride and ferroelectric research communities. While the ferroelectric community had established a clear understanding of spontaneous polarization,\cite{resta2007theory} this separation allowed an incorrect definition of spontaneous polarization to persist within the nitride community for many years.\cite{Bernardini}

Therefore, as in the work by Dreyer et al.\cite{Dreyer2016}, the most logical reference for wurtzite is a layered hexagonal structure, which is centrosymmetric, nonpolar, and has vanishing formal polarization (modulo $P_q$). At first glance, this choice may seem just convenient but unconventional, as there is no experimental evidence of such a structure for InN, GaN or AlN. However, the layered hexagonal structure serves an intermediate point in the ferroelectric switching for wurtzite, directly linking it to experimental ferroelectric measurements of polarization.\cite{Mo2024} This approach allows spontaneous polarization to be unequivocally defined and connected to the ferroelectric switching through the formal polarization difference between the layered hexagonal and wurtzite structures. In general, spontaneous polarization is the difference in the formal polarization of a polar crystal and its non-polar centrosymmetric structure associated through ferroelectric switching.\cite{Resta1994,resta2007theory} Analogously, the spontaneous polarization is equal to half of the formal polarization difference between the two opposite polarities of a polar structure. This definition of spontaneous polarization is advantageous: First, it is directly linked to experimental polarization measurements through ferroelectric switching. Second, the choice of reference is not arbitrary — it is always a centrosymmetric, nonpolar structure that serves as the midpoint between the switching of polarities.  It is also consistent with the etymological origin of the concept of spontaneous polarization within ferroelectric materials.\cite{kanzig1957ferroelectrics,resta2007theory} Note that because the layered hexagonal formal polarization is zero (modulo $P_q$), the value of the formal polarization of wurtzite is the same as the spontaneous polarization. Therefore, for wurtzite, \textit{spontaneous polarization concides with the formal polarization of a material at zero strain}.\cite{Ambacher1999,Mo2024}

\begin{figure*}

\includegraphics[width=17.5cm]{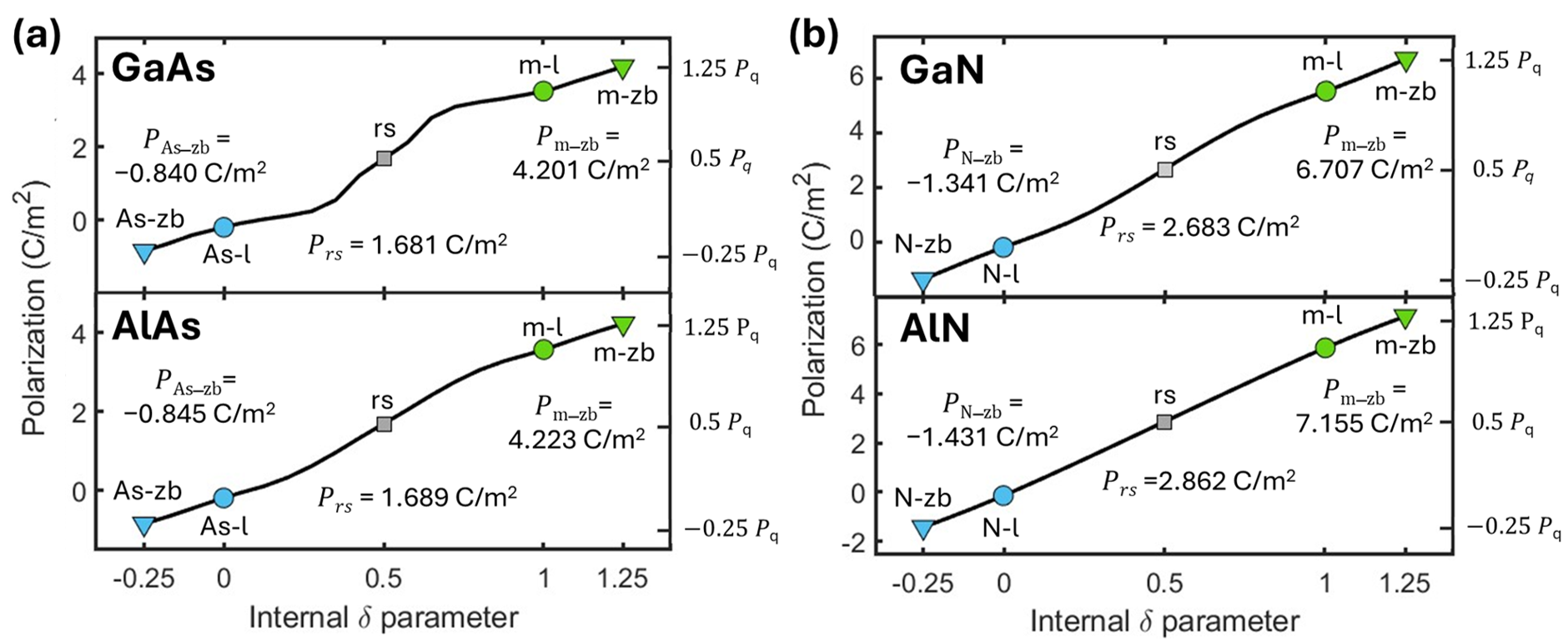}

\caption{Formal polarization calculated for the structures under the zincblende N-polar to m-polar transformation for (a) GaAs and AlAs and for (b) GaN and AlN for a branch with $P_f^{rs}=0.5 P_q$.}
\label{PNA}
\end{figure*}

As discussed earlier, the spontaneous polarization of a zincblende crystal has always been considered zero, based on the argument that its four polar triad axes cancel out any net spontaneous polarization.\cite{Nye1985} However, the <111> direction of zincblende is polar, has a nonzero formal polarization, and also exhibits piezoelectricity.\cite{piezoingaas} This system is of extreme technological importance, and there is abundant experimental data for its interfaces; however, there is no experimental report of ferroelectric switching of zincblende. To understand the spontaneous polarization of zincblende crystals, we can devise a transformation, analogous to the wurtzite system, that switches polarities through its centrosymmetric, nonpolar structure. This transformation is illustrated in figure \ref{schem}, exemplified for the III-N material system. N-polar zincblende is transformed by displacing metal atoms in the <111> direction, resulting in an N-polar layered structure (distinct from a layered hexagonal structure), and then further transforming into rocksalt, its centrosymmetric counterpart. The transformation then continues through a m-polar layered phase before reaching m-polar zincblende. Note that we adopt the conventional wurtzite nitride terminology to refer to N-polar zincblende, which, in the zincblende community, is commonly described as (-1-1-1) oriented zincblende with a (111)B surface or a nitrogen-terminated (111) surface.\cite{zincblendefaces} This surface is stable only when terminated with nitrogen. However, polarity and surface termination are distinct concepts, even though, in practice, N-polar zincblende exhibits a nitrogen-terminated surface. The illustrated transformation is an extension of the one in this work by Adamski et al.\cite{Adamski2019} to connect rocksalt and zincblende. Through this transformation representing the ferroelectric switching of zincblende, we can define its spontaneous polarization as the formal polarization difference relative to rocksalt.

This system, however, exhibits more subtleties than its wurtzite counterpart. Notably, the formal polarization of its centrosymmetric structure (rocksalt) is nonzero. It is impossible to construct a stoichiometric rocksalt primitive unit cell that has no dipole along the <111> direction; in fact, the minimum absolute value of formal polarization is $P_q/2$.\cite{Adamski2019,Spaldin-halfpq} However, in finite crystals, such large polarizations are not observed experimentally, as they are mitigated by external species, incomplete surface coverage, or rough, nonpolar surface reconstructions.\cite{Rocksalt-polarization,Surface-polarization} Consequently, in the case of zincblende, the spontaneous polarization is referenced to the nonzero formal polarization of rocksalt. As a result, unlike in the wurtzite system, the spontaneous polarization value is not present in the formal polarization lattice.

We have calculated the formal polarization along the zincblende N-polar to m-polar transformation shown in \ref{schem} using the PWscf code of the Quantum Espresso (QE) software package.\cite{QE-2009,QE-2017} The unit cell consisted of 6 atoms for all the materials considered. We used the generalized gradient approximation Perdew-Burke-Ernzerhof (PBE) functional for exchange correlation. The wave functions of the valence electrons are represented by a plane-wave basis set with a cutoff energy of 70 Ry and the electron density and effective Kohn-Sham potential by discrete Fourier series with a cutoff energy of 560 Ry. The interactions of valence electrons with the atomic nuclei and core electrons are described by pseudopotentials taken from the open-source Standard Solid State Pseudopotentials (SSSP PBE Efficiency v1.3.0) library.\cite{SSSP} Brillouin zone integrals were evaluated on a Monkhorst-Pack mesh of 15 × 15 × 30 kpoints. The convergence threshold was set to $10^{-8}$ Ry for the total energy and to $10^{-6}$ Ry/Bo for the forces on atoms. We have verified that the crystals preserved a bandgap throughout the transformation.

In figure \ref{PNA}, we show the formal polarization calculated for the structures under the ferroelectric switch of zincblende for (a) GaAs and AlAs and (b) GaN and AlN in the branch showing $P_f^{rs}=0.5 P_q$. The $\delta$ parameter is defined as the ratio of the separation
between planes of anions and cations to the separation between planes of cations.\cite{Adamski2019} The spontaneous polarization deduced from the formal polarization difference between zincblende and rocksalt is smaller for arsenides ($P_{sp}^{GaAs}=2.521$ $ C/m^2$ and $P_{sp}^{AlAs}=2.534 $ $C/m^2$) than for nitrides ($P_{sp}^{GaN}=4.024$ $ C/m^2$ and $P_{sp}^{AlN}=4.293$ $C/m^2$) due to their larger unit cells. The consistent definition of spontaneous polarization enables us to demonstrate that, contrary to historical belief, zincblende does indeed have a spontaneous polarization. Remarkably, the spontaneous polarization of zincblende is about three times larger than its wurtzite counterpart for each of the materials. Additionally, this is an interesting example to show the fundamental difference between the spontaneous and the formal polarization. Spontaneous polarization is strictly related to ferroelectric properties and requires a reference structure, as the physical mechanism of ferroelectricity involves switching between two structures with different polarities. On the other hand, formal polarization can be used to predict charge gradients between epitaxially grown crystals as described in the interface theorem by Vanderbilt.\cite{interfacetheorem}

\begin{figure*}
\includegraphics[width=17cm]{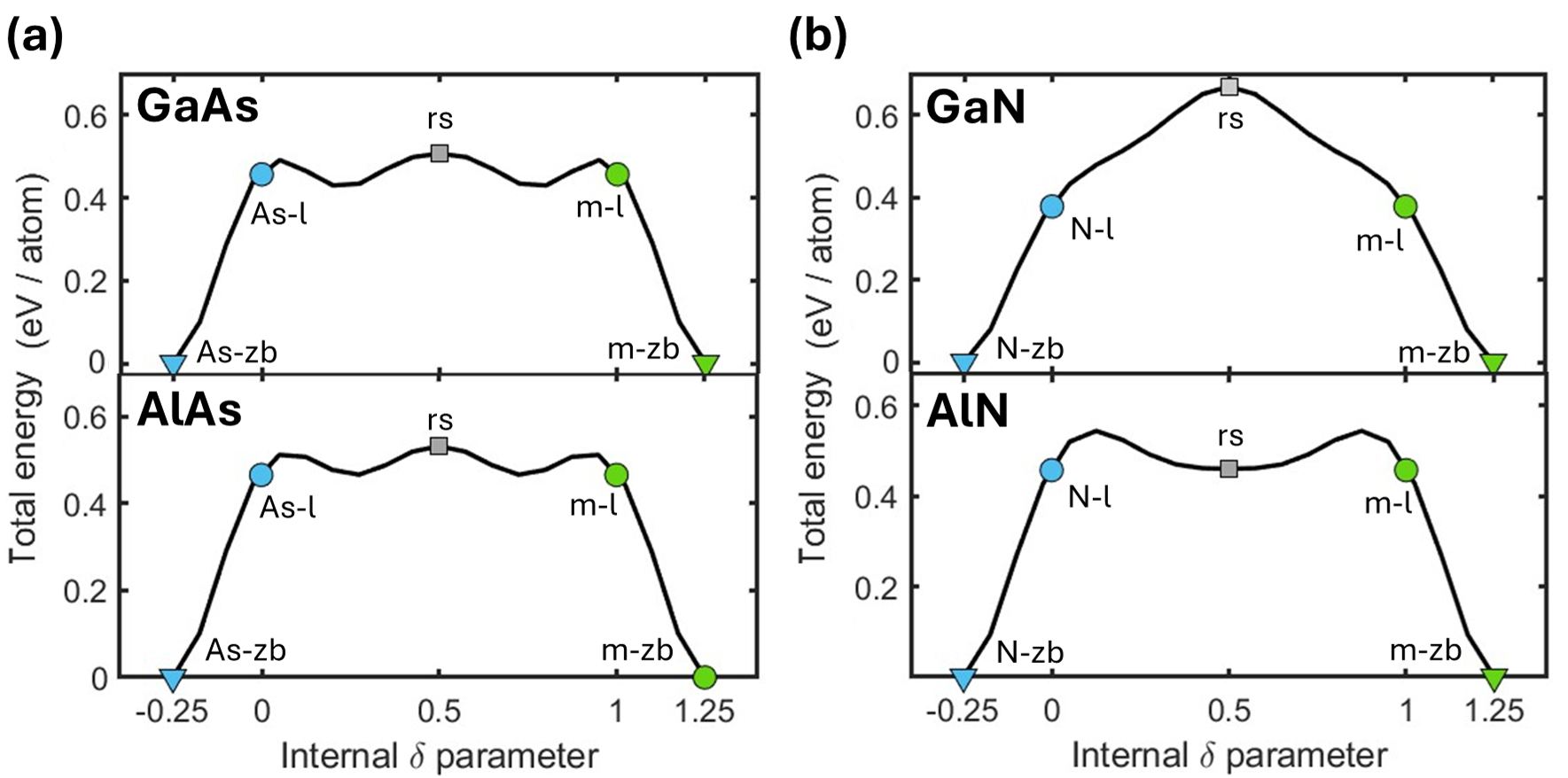}

\caption{Energy per atom calculated for the structures under the zincblende N-polar to m-polar transformation for a) GaAs and AlAs and b) GaN and AlN.}
\label{ENA}
\end{figure*}

Interestingly, the formal polarization of zincblende always takes the same value in III-V materials, expressed in quantum of polarization units ($P_q$), as it does for rocksalt ($P_f^{rs}$=0.5 $P_q$). For zincblende, the formal polarization is always -0.25 $P_q$ for the group-V polar structure and 1.25 $P_q$ for the group-III polar structure. This implies that the formal polarizations of zincblende (and rocksalt) do not explicitly depend on the atoms involved in bonding, but only on the volume of the unit cell $(\Omega)$ via $P_q$. More precisely, the quantum of polarization for this transformation depends only on the basal plane area ($A_b$) of the unit cell: $P_q= e/A_b$. 
The ionic contribution to the formal polarization will always be the same, as the valence of the elements is the same for III-V materials, and only the ionic contribution from electrons involved in bonding is considered (the rest of the nucleus charge is screened). While the wavefunctions describing the electronic bonding between different nitrides and arsenides are indeed different, the symmetry of the zincblende and rocksalt lattices ensures that the electronic part of the polarization always cancels out. As for rocksalt, no large surface charge is observed (111)-oriented freestanding zincblende crystals, due to its non-vanishing formal polarization, as a result of its passivation with external species or partial coverage.\cite{zincblende-polarization-pasivation,polarsurfaceszincblende}

There is no experimental evidence of ferroelectric switching in zincblende structures. However, piezoelectric and pyroelectric-like effects have been observed experimentally\cite{piro-like} resulting from changes in the $\delta$ parameter. This means that when the zincblende structure is uni-axially deformed in the <111> direction, the cation-to-anion distance changes, deviating from that in the ideal zincblende structure. This leads to a non-zero contribution from the electronic part of the polarization, resulting in piezoelectricity and a formal polarization $P_f \neq 0.25 P_q$.  In fact, this occurs whenever a zincblende A layer is epitaxially grown on top of another zincblende B. The A layer adopts the in-plane lattice parameter of B and deforms in the <111> direction, modifying its $\delta$. The formal polarization of the B layer is $P_f^b = 0.25 P_q^b$, while the epitaxial A layer has a formal polarization $P_f^a = (0.25+\alpha) P_q^b$, where $\alpha$ is the change of the formal polarization due to the non-ideal $\delta$ parameter. While in the case of GaAs/AlAs interfaces, the lattice mismatch is small enough that this effect ($\alpha$) is negligible, interface charges between lattice-mismatched zincblende structures are well explained by this phenomenon.\cite{piezoingaas}

The reason ferroelectric switching has never been observed experimentally in zincblende is due to the large energy barrier for switching and the associated high coercive electric field. In figure \ref{ENA}, we show the energies per atom calculated for the structures under the ferroelectric switch of zincblende for (a) GaAs and AlAs and (b) GaN and AlN. The curves for the arsenides show two local maxima at similar energies, one near the layered structure and the other for rocksalt. AlN also displays its maximum energy configuration near the layered structure, but rocksalt is a local minimum. For GaN, the maximum energy is located at rocksalt, significantly higher than that for the layered structures. Specifically, for arsenides and aluminum nitride, the energy barrier is about 0.5 eV/atom, while for GaN, it exceeds 0.6 eV/atom. These barriers are roughly double the energy required to switch conventional binary nitrides wurtzite structures, which already requires very large coercive fields. For instance, coercive fields are experimentally over the critical electric field ($\sim$10 MV/cm) for wurtzite AlN. This is why its ferroelectric properties were not experimentally observed until recently, when alloying it with ScN helped lower the coercive field \cite{Fichtner2019}. Indeed, the estimated coercive fields for the studied zincblende materials, based on our calculations, exceed 40 MV/cm, making the switching of binary nitrides or arsenides impossible, as their critical fields are much lower. Furthermore, while the zincblende phase has a bandgap close to that of wurtzite, the rocksalt phase has a smaller bandgap and thus a smaller critical electric field, making it even more difficult to sustain the coercive field without breakdown. As shown, even if the transformation exists, its switching barrier is too high for the material to sustain the transformation. However, as with wurtzite, group-III transition metal ternary alloys can lower the rocksalt energy.\cite{wang-barrier-scaln} It is intriguing whether it is possible to find a material candidate with a low enough coercive field for switching zincblende. However, it remains unclear whether alloys with such transition metals would lower the energy for the layered structures, as they may still present the maximum energy barrier for the ferroelectric switching of zincblende. Nonetheless, the definition of spontaneous polarization in this phase is unambiguous and rigorous. Even though the ferroelectric switch of this phase is hardly achievable experimentally, the spontaneous polarization value is well-defined as half the formal polarization value between its polar phases. Alternatively, a potential strategy to demonstrate ferroelectric switching in zincblende could involve inducing it through proximity switching of a ferroelectric wurtzite nitride.\cite{Skidmore2025-proximityferro}

In conclusion, we have discussed the definition of spontaneous polarization and clarified that spontaneous polarization is strictly a bulk property that affects ferroelectric switching. This definition allows for a unique reference structure, the non-polar centrosymmetric associated structure, to be used for calculating the spontaneous polarization, thereby eliminating ambiguities. Extending this definition to the zincblende case, we identified a transformation along the (111)-direction, switching zincblende N-polar to m-polar through the non-polar centrosymmetric rocksalt structure. Using this transformation, we have shown that the spontaneous polarization of zincblende is about three times higher than that of wurtzite, contrary to the historical belief that it is zero. We have explained how this fact is compatible with the experimental observations of piezoelectricity, interface bound-charges and absence of polarization effects in freestanding crystals of zincblende. The ferroelectric coercive field of zincblende materials is too large compared to their typical critical electric fields, preventing ferroelectric switching in the studied materials, which explains the absence of experimental evidence. 
\\

The first two authors contributed equally to this manuscript.
This project has received funding from the European Union under the Marie Skłodowska-Curie grant agreement No 101146464 and from the German Science Foundation (DFG-project: "Polrock" 530081697).

The data that support the findings of this study are available from the corresponding author upon 
 reasonable request.

\bibliography{aipsamp}

\end{document}